\documentclass[%
 aip,
 amsmath,amssymb,
 reprint,%
]{revtex4-1}
\usepackage{amsmath}
\usepackage{multirow}
\usepackage{hhline}
\usepackage{graphicx}
\usepackage{dcolumn}
\usepackage{bm}

\usepackage[utf8]{inputenc}
\usepackage[T1]{fontenc}
\usepackage{mathptmx}
\usepackage{etoolbox}

\makeatletter
\def\@email#1#2{%
 \endgroup
 \patchcmd{\titleblock@produce}
  {\frontmatter@RRAPformat}
  {\frontmatter@RRAPformat{\produce@RRAP{*#1\href{mailto:#2}{#2}}}\frontmatter@RRAPformat}
  {}{}
}%
\makeatother
\begin{document}

\preprint{AIP/123-QED}
\title{Polarized Raman Analysis at Low Temperature to Examine Interface Phonons in $InAs/GaAs_{1-x}Sb_x$ Quantum Dot Heterostructures\\ }

\author{Priyesh Kumar}
\affiliation{ 
Indian Institute of Technology  Gandhinagar, Gandhinagar-382055, Gujarat, India. 
}%
\author{Sudip Kumar Deb}
\affiliation{%
Centre for Research in Nanotechnology \& Science, Indian Institute of Technology Bombay, Mumbai-400076, Maharashtra, India
}%
\author{Subhananda Chakrabarti}
\affiliation{%
Department of Electrical Engineering, Indian Institute of Technology Bombay, Mumbai-400076, Maharashtra, India
}%

\author{Jhuma Saha}
 \email{jhuma.saha@iitgn.ac.in}
\affiliation{ 
Indian Institute of Technology  Gandhinagar, Gandhinagar-382055, Gujarat, India. 
}%

\date{\today}

\begin{abstract}
An experimental study of optical phonon modes, both normal and interface (IF) phonons, in bilayer strain-coupled $InAs/GaAs_{1-x}Sb_x$ quantum dot heterostructures has been presented by means of low-temperature polarized Raman scattering. The effect of $Sb$-content on the frequency positions of these phonon modes has been very well correlated with the simulated strain. The Raman peaks show different frequency shifts in the heterostructure with varying $Sb$-content in the capping layer. This shift is attributed to the strain relaxation, bigger size of quantum dots and type-II band alignment. 
\end{abstract}

\maketitle

\section{\label{sec:level1}Introduction\protect}
Self-assembled $InAs$ based quantum dot (QD) devices have garnered significant attention from the research community due to their potential for use in a broad spectrum of optoelectronic applications. These self-assembled QDs offer the advantage of creating defect-free systems that confine charge carriers within nanometer-scale dimensions without the need for complex lithography and etching processes. These QDs are usually produced by the Stranski-Krastanov process during the growth of mismatched materials using molecular beam epitaxy (MBE). In this process, interlayer strain is essential in creating pyramid-shaped or hemispherical dots on a thin wetting layer. The strain-coupling in these self-assembled QD systems is particularly advantageous as it facilitates the vertical ordering of QDs in multilayer QD systems. This vertical alignment increases the volume of active regions, thereby enhancing absorption efficiency.  \cite{tutu2012improved, kim2004high, panda2018optimization} Moreover, the linewidth of photoluminescence (PL) spectra and the uniformity in dot size can be precisely controlled, making these systems highly desirable for various applications. \cite{panda2017evidence} Specifically, this strain-coupling effect has been effectively utilized in $InAs/GaAs$ QD systems to achive an emission wavelength of $1.3\mu m$, which is highly relevant for telecommunications applications. \cite{dowd19991, ledentsov2002applications} Given the significance of these strain-coupled heterostructures, they have become a focal point of research in device applications, particularly in the telecommunications sector. However, there is ongoing research to push the emission wavelength beyond $1.3\mu m$ to meet the increasing demands of advanced telecommunications technologies. One promising approach to achieving this is the incorporation of $Sb$ into the the system, where $GaAsSb$ is used as a capping layer in $InAs/GaAs$ QD systems. This modification increases the aspect ratio of the QDs, thereby reducing the strain within them and leads to a redshift of the emission wavelength. The $InAs/GaAs_{1-x}Sb_x$ QD systems have gained popularity due to the intriguing properties that arise from varying the $Sb$ composition. Notably, it has been observed that the band alignment in these systems shifts from type-I to type-II beyond a certain  $Sb$-composition (approximately $14-16\%$).\cite{ulloa2012analysis, ripalda2005room, jang2008carrier} This unique property allows $InAs/GaAs_{1-x}Sb_x$ QDs to facilitates emission wavelength beyond $1.5\mu m$, which is particularly valuable for advanced telecommunications. \cite{akahane20061} For instance, recent work by Saha et al\cite{saha2019broad} reported a room temperature emission at approximately $\sim1.7\mu m$ from the strain-coupled bilayer $InAs/GaAs_{1-x}Sb_x$ QD heterostructures, further highlighting the potential of these systems for next-generation optoelectronic devices. 

The generation of quantum dots (QDs) is heavily dependent on strain, which has important ramifications for QD applications in various devices. Understanding how strain affects the electronic, vibrational, and optical properties of QDs is therefore essential. Numerous studies have been conducted to explore the impact of strain on the electronic and vibrational (phonon) structures of QDs using techniques such as X-ray diffraction, Raman spectroscopy, photoluminescence (PL), and photoluminescence excitation spectroscopy. Phonons are essential to electron-phonon scattering, which affects the carrier relaxation process, which is a crucial element in determining the performance of semiconductor devices. For this reason, the study of phonon characteristics is especially relevant to device applications. Raman spectroscopy is an effective tool for investigating phonon modes, as it provides valuable insights into doping levels, crystalline quality, impurity concentrations, and other material characteristics. Despite its advantages, the study of phonon properties in strained QD systems using Raman spectroscopy has been limited. This is primarily because the Raman signals from QDs are often weak due to the small scattering volume of the QDs compared to the surrounding capping material and substrate. Additionally, self-assembled QDs typically exhibit strongly corrugated interfaces, which can lead to the emergence of IF phonon modes that are localized at the tips and cusps of the QDs, as predicted by Knipp and Reinecke\cite{knipp1992classical} using the dielectric continuum model. There have been several studies focusing on the IF phonon modes in QDs, utilizing techniques such as PL and both normal and resonance Raman spectroscopy. These studies have successfully identified various IF modes and examined their behavior in response to different growth parameters. For example, Puesp et al\cite{pusep1998raman} reported the presence of IF modes localized near the edges of the $InAs$ QDs in $GaAs$ and in ($In,Ga,Al$)$Sb/GaAs$ structures. In further research, Puesp et al\cite{pusep1998raman} used resonance Raman scattering at the $E_0+\Delta_0$ gaps of $InAs$ and $GaAs$ to identify distinct IF modes and demonstrated a strong influence of strain on these modes. These findings underscore the importance of studying phonon properties in strained QD systems, as they provide critical insights about how strain affects the electronic and vibrational properties of the material.

Despite the significant interest in the optical and electronic properties of $InAs$/$GaAs_{1-x}Sb_x$ quantum dot (QD) systems, there has been a noticeable lack of studies focusing on the normal and IF phonon modes in these structures. The work by  Dai et al\cite{dai2015raman} represents one of the few exceptions, where they examined $InAs$ QDs capped with $GaAs_{1-x}Sb_x$ using an $Sb$ spray technique applied immediately prior to $GaAs$ capping. Their research highlighted changes in photoluminescence (PL) emission intensity and a redshift in emission wavelength corresponding to different durations of $Sb$ spray. However, the weak signals obtained in their study limited the unambiguous identification of distinct phonon modes. Dai et al\cite{dai2015raman} concluded that the $Sb$ spray likely reduced defects and contributed to strain relaxation by facilitating the formation of $GaAsSb$ at the QD/cap interface. In this paper, we address this gap by presenting the results and analysis of our low-temperature polarized Raman measurements aimed at investigating the frequency positions of both normal and IF phonons in $InAs/GaAs_{1-x}Sb_x$ bilayer QDs heterostructures. Our study seeks to provide a clearer understanding of the phonon modes in these strained systems, which is critical for advancing the development of QD-based optoelectronic devices.
\begin{figure}
\includegraphics[width=0.29\textwidth]{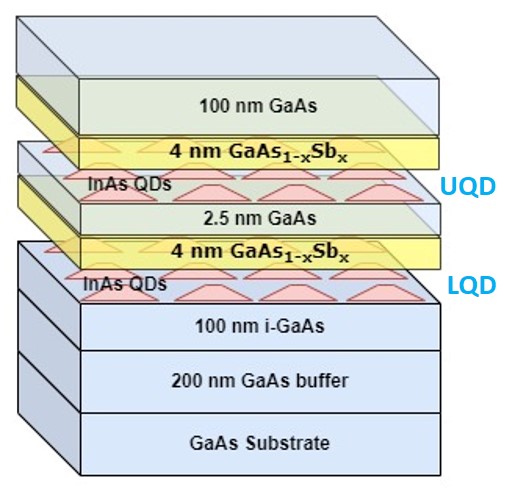}
\caption{\label{fig:Schematic_QD}Schematic of bilayer QD heterostructure .}
\end{figure}

\section{Experimental details}
\begin{figure}
\includegraphics[width=1
\columnwidth]{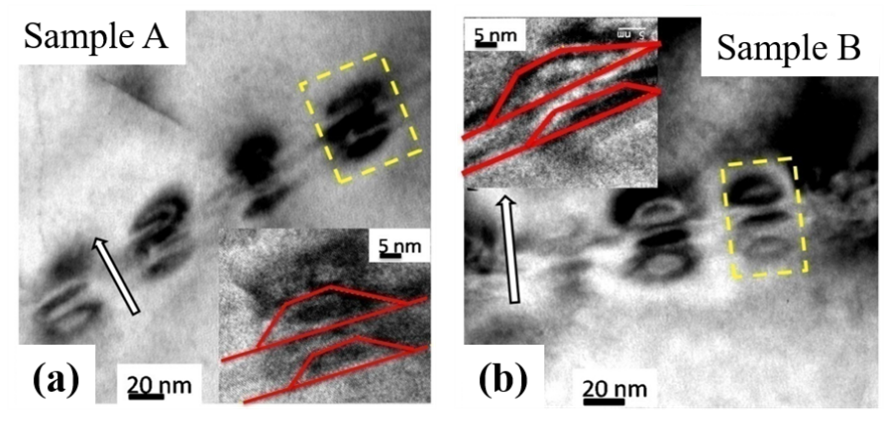}
\caption{\label{fig:TEM_QD} Cros-sectional TEM images of (a) Sample A $\&$ (b) Sample B.}
\end{figure}

Strain-coupled bilayer $InAs/GaAs_{1-x}Sb_x$ QDs were grown by the MBE system via Stransky-Krastanov growth. Fig.~\ref{fig:Schematic_QD}  illustrates the schematic representation of the developed heterostructure, featuring two layers of $InAs$ QDs capped with a $GaAs_{1-x}Sb_x$ capping layer. The bilayer QD architecture is characterized by strain coupling with a $GaAs_{1-x}Sb_x/GaAs$ spacer, wherein the lower QD possesses a $2.5 ML$ coverage, while the upper QD (UQD) has a 3.2 ML coverage. Consequently, the UQD exhibits larger dimensions compared to the lower QD (LQD), as evidenced by a Transmission Electron Microscopy (TEM) image as shown in Fig.~\ref{fig:TEM_QD}. The two dot layers are separated by a spacer layer of $6.5 nm$ and capped with $4 nm$ $GaAs_{1-x}Sb_x$ capping material. The $Sb$-composition has been taken as $10$\% and $20$\% in the capping material and referred as samples A and B, respectively. The details of the growth have been mentioned in our previous study. \cite{saha2019broad} 
\begin{figure}
\includegraphics[width=1
\columnwidth]{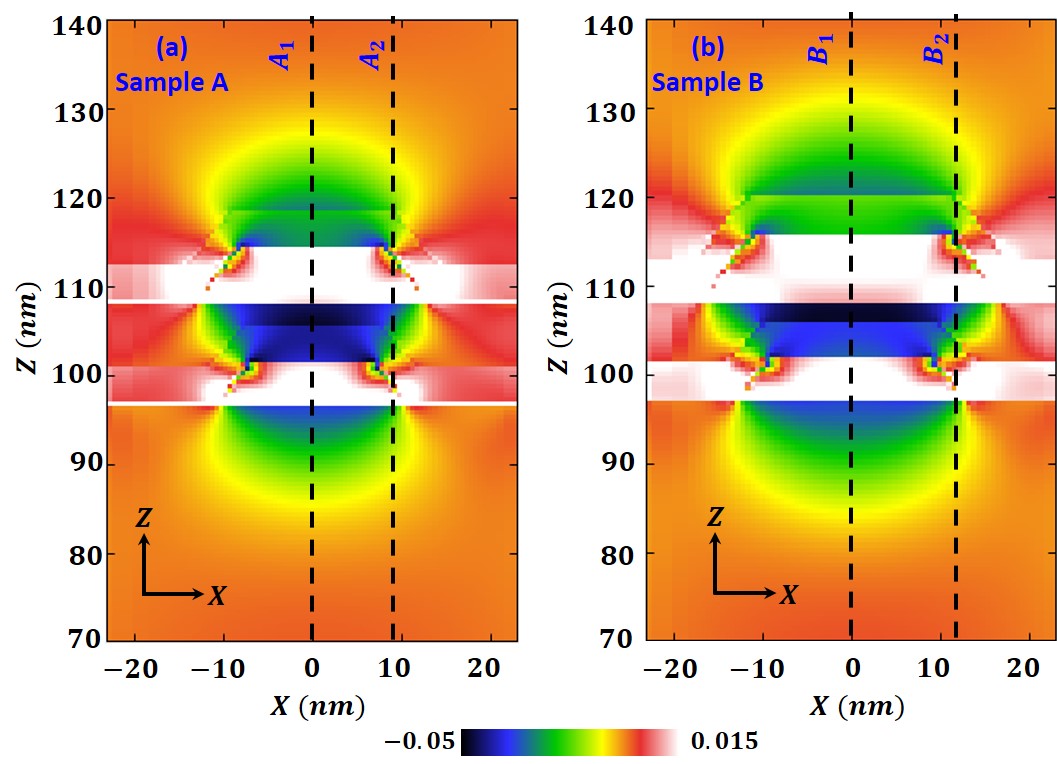}
\caption{\label{fig:Nextnano_QD_contour}Strain field contour in (a) sample A, $\&$  (b) sample B. The vertical dash lines indicate cutlines at which the strain profile has been observed. Cutline 1 ($A_1$ and $B_1$ for sample A $\&$ sample B respectively) is through the center of QD and cutline 2 ($A_2$ and $B_2$ for sample A $\&$ sample B respectively) is through the edge of QD.}
\end{figure}
\begin{figure*}[!ht]
    \centering
    \includegraphics[width=0.95\textwidth]{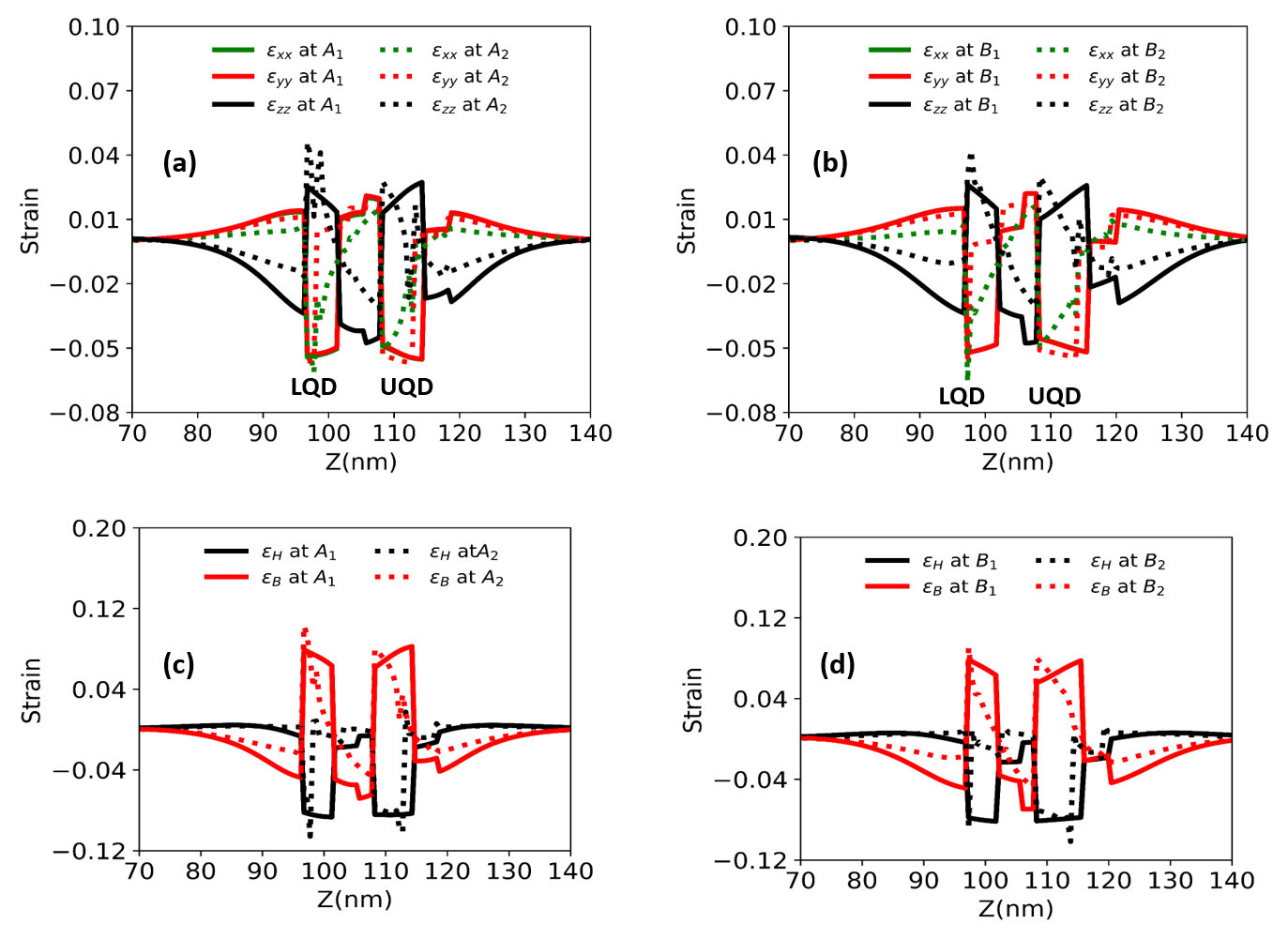}
    \caption{\label{fig:Strain} Strain profiles of (a) sample A at cutlines $A_1$ and $A_2$ $\&$ (b) sample B at cutlines $B_1$ and $B_2$. Biaxial ($\varepsilon_B$) and hydrostatic strain ($\varepsilon_H$) profiles of (c) sample A at cutlines $A_1$ and $A_2$ $\&$ (d) sample B at cutlines $B_1$ and $B_2$.}
\end{figure*}
A $532 nm$ frequency doubled Nd:Yag laser focused on a $2-3\mu m$ spot with $10 mW$ of power at the laser head was used to excite the Raman spectra. Spectra were collected in backscattering geometry and analyzed using an HR800-UV confocal micro-Raman spectrometer. The measurements were acquired in $z(yx)\overline{z}$ and $z(xx)\overline{z}$ polarization with $x\Vert[110], y\Vert[1\overline{1}0]$  and $z\Vert[001]$. Samples were cooled to $77K$ to reduce the width of the different Raman lines so that the very weak IF phonon modes could be observed in the presence of comparatively stronger normal phonon modes. The heterostructures were simulated by Nextnano software\cite{birner2007nextnano} to visualize the distribution of strain field in the different layers and dimensions of QD were taken according to the TEM image shown in Fig.~\ref{fig:TEM_QD}. This software comprises a complete database of group III-V materials' characteristics and allows for 3-dimensional simulations of the quantum mechanical electronic structure within the heterostructure. We have already mentioned the software related details in our previous work by Saha et al.\cite{saha2019broad}

Fig.~\ref{fig:Nextnano_QD_contour} shows the contour of the strain field in both samples, wherein both QDs are under tensile strain. Generally, tensile strain is positive, and compressive strain is negative in magnitude. The strain is compressive at the bottom of LQD, tensile inside the QD region, compressive towards the capping layers, and finally tensile at the end of the capping layer of UQD. This can be attributed to the larger lattice constant of InAs compared to GaAs. Additionally, it can be observed that the strain field is different at the center and the edge of QD. Two cutlines, cutline $1$ ($A_1$ and $B_1$) and cutline $2$ ($A_2$ and $B_2$), are marked at the center and edge of QD respectively, and these markings are used to analyze strain later in the manuscript. The magnitude of strain in sample B is lower as compared to sample A. This is due to the bigger size of QDs\cite{saha2019broad} and the presence of high $Sb$-composition in the capping layer of sample B, which reduces the lattice mismatch between the InAs QDs and $GaAs_{0.8}Sb_{0.2}$ capping layer, and thereby decreasing the strain within the entire heterostructure. 

\section{Theoretical Analysis}
\begin{figure*}[!ht]
\includegraphics[width=1\textwidth]{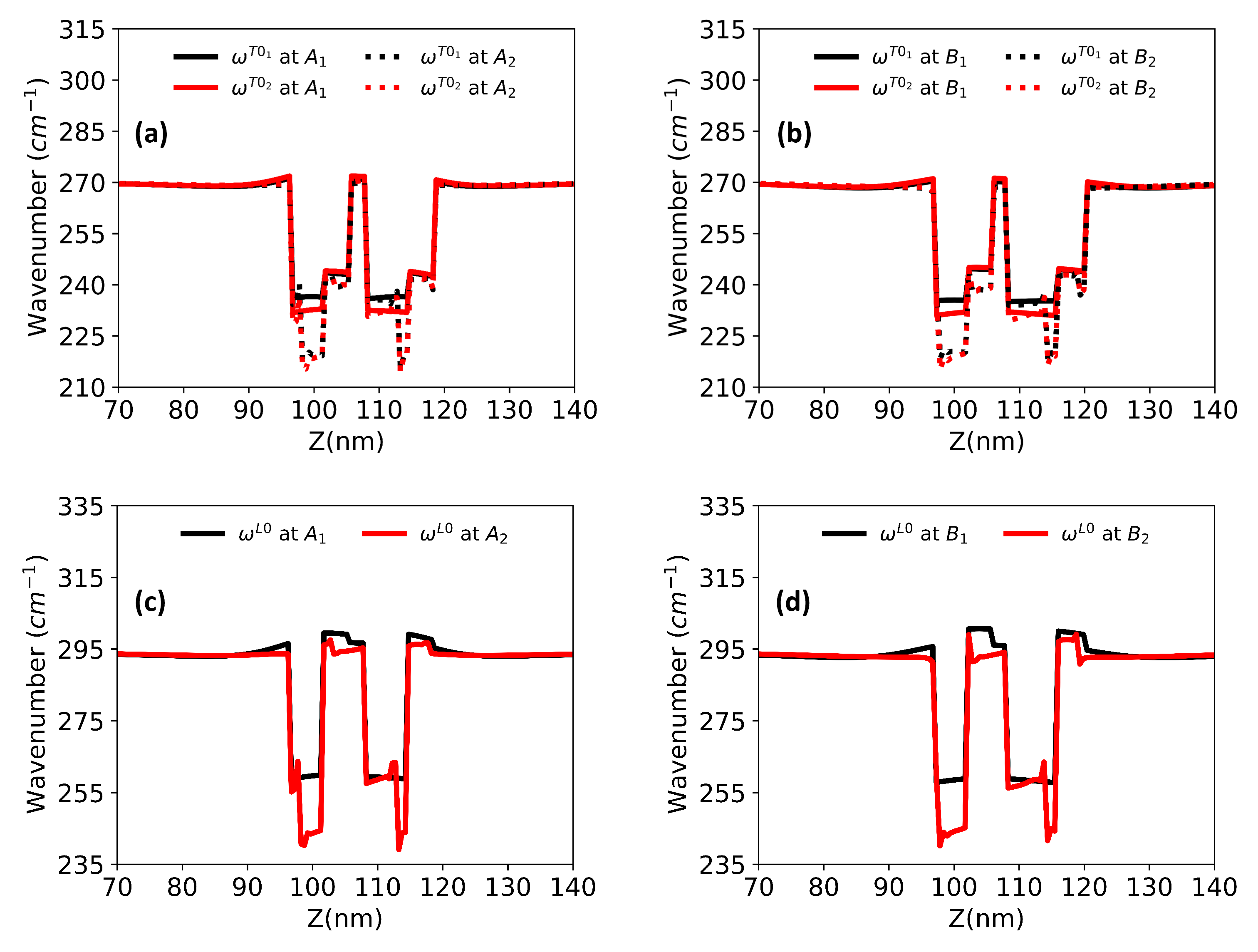}
\caption{\label{fig:my_plot}Shifted $TO$ phonons of (a) sample A at cutlines $A_1$ and $A_2$ $\&$ (b) sample B at cutlines $B_1$ and $B_2$. Shifted $LO$ phonons (c) sample A at cutlines $A_1$ and $A_2$ $\&$ (d) sample B at cutlines $B_1$ and $B_2$.}
\end{figure*}
In order to assign the different phonon modes, it is necessary to analyze and estimate quantitatively the effects of strain on different phonon modes. As shown by the simulation of strain profiles in both the samples, both UQD and LQD are under considerable strain and it is known that the phonon mode frequencies exhibit both blue-shift as well as red-shift, by compressive and tensile strains. Thus, it is necessary to estimate the strain profiles over the QD heterostructures to provide explanation for the origin of the different Raman lines in both samples. Cardeira et al \cite{cerdeira1972stress} have carried out a detailed examination of stress-induced Raman mode shifts in various zinc-blende and diamond semiconductors and we use their framework to estimate the position of different phonon modes.
The strain tensor components can be broadly expressed as hydrostatic and biaxial strain denoted by $\varepsilon_H$ and $\varepsilon_B$ respectively. These are defined as\cite{xiong2017accurate}

\begin{equation}
\varepsilon_H =\varepsilon_{xx}+ \varepsilon_{yy}+ \varepsilon_{zz}
\label{eq:one}
\end{equation}
\vspace{-3.5mm}
\begin{equation}
\varepsilon_B =\varepsilon_{zz}- 0.5\times{(\varepsilon_{xx}+\varepsilon_{yy})}
\label{eq:two}
\end{equation}


Where, $\epsilon_{ij}$ is the diagonal components of the strain tensor. Fig.~\ref{fig:Strain} (a) and (b) for samples A and B, respectively, at both cutlines, display the simulated strain components. The $\varepsilon_{xx}$ and $\varepsilon_{yy}$ strain profiles are similar (compressive), as the definition of material layers are the same in the both x and y directions, whereas the strain is tensile in the z-direction. The changes in the phonon frequencies for the transverse optical modes ($TO$) and longitudinal optical mode ($LO$) can be calculated following the prescription outlined by Cardeira et al\cite{cerdeira1972stress} where they have shown that the application of uniaxial stress causes a shift of the $LO$ phonon, while the $TO$ mode shows both a splitting and shift. They had further shown that the shifts can be expressed as due to hydrostatic and biaxial stain ( $\varepsilon_H$ and $\varepsilon_B$ respectively ) and can be calculated in terms of the material properties  $p$, $q$, and $r$ which outline the modifications to the “spring constants” of the $k=0$ optical phonons. These parameters have been listed by Cardeira et al\cite{cerdeira1972stress} for a set of III- V semiconductors. The unperturbed $LO/TO$ phonons are denoted by $\omega_{L0}/\omega_{T0}$ and the shifted ones are by $\omega^{L0}/\omega^{T0}$ respectively. The frequency shift of the $TO$ and $LO$ modes as a function of hydrostatic and biaxial strain along the growth direction of the QD heterostructures can be used to calculate and interpret the various observed lines in the Raman spectra for the two samples (discussed in next section). We note here that $GaAs_{1-x}Sb_x$ exhibits two mode behavior with $GaAs$ like $LO/TO$ and $GaSb$ like $LO/TO$ modes with $GaAs$ like $LO$ modes are most intense\cite{yano1989molecular}. For $x \approx 10\%$, the $GaAs$ like $LO$ mode for $GaAs_{1-x}Sb_x$ is more or less the same as that for $GaAs$ $LO$  and the $TO$ mode\cite{yano1989molecular} is at $238 cm^{-1}$.  However, for $x \approx 20\%$, we use $LO$ phonon mode\cite{yano1989molecular} at $285 cm^{-1}$ and for $TO$ phonon $240cm^{-1}$. The frequencies of $GaAs$, $InAs$, and $GaAsSb$ layers in the lower and upper QDs and the capping layers have been calculated and can be Visualized in terms of depth. It is evident that the strain profile is slightly different at the center ($A_1$ and $B_1$) and at the edge ($A_2$ and $B_2$) and the Raman frequencies have also been calculated for these positions. 
Fig.~\ref{fig:Strain} (c) and (d) show $\varepsilon_H$ and $\varepsilon_B$ for both of the samples A and B respectively Corresponding to growth direction covering both the UQD and LQD. It can seen that the variation of both $\varepsilon_H$ and $\varepsilon_B$ over LQD and UQD at the center of QD ($A_1$ and $B_1$) and edge of QD ($A_2$ and $B_2$) are quite different for both of the samples. In particular, $\varepsilon_B$ over LQD has a sharp spike while its variation over UQD is more gradual. Using Eqs. 39 and 42 from MS thesis of Hussey\cite{hussey2007raman} and $\varepsilon_H$/$\varepsilon_B$ strain distribution we can determine the phonon shift for the $TO$ and $LO$ modes, as shown in Fig.~\ref{fig:my_plot}. The figure depicts the shift in $LO$ and $TO$ phonon modes in the UQD region for both samples. It is mainly because most of the Raman signal originates from top $\approx 100nm$. Since $LO$ modes are usually stronger compared to $TO$ modes we concentrate on Fig.~\ref{fig:my_plot} (c) and (d). The $LO$ phonon over most of the region around the center of QD ($A_1$ and $B_1$) is $\approx258-260$ $cm^{-1}$ with a sharp downward spike at edge of QD ($A_2$ and $B_2$). This is to be mentioned here that there are different expressions for the biaxial strain as a function of different strain components $\epsilon_{ij}$ used by different authors.
\begin{equation}
\varepsilon_B = \sqrt{(\varepsilon_{xx} - \varepsilon_{yy})^2 + (\varepsilon_{yy} - \varepsilon_{zz})^2 + (\varepsilon_{zz} - \varepsilon_{xx})^2}
\label{eq:three}
\end{equation}
\vspace{-3.5mm}
\begin{equation}
\varepsilon_B = 2\varepsilon_{zz} - (\varepsilon_{xx} + \varepsilon_{yy})
\label{eq:two}
\end{equation}


To compare the magnitude of shifts obtained using the different expressions, we have also performed calculations using two other expressions ( Eqs.~(\ref{eq:three})\cite{cusack1996electronic}  and  Eqs.~(\ref{eq:four})\cite{grundmann1995inas}) for biaxial strains. However, we observe that the changes are very small and it can be attributed to the fact that the biaxial strain for these materials ($InAs$, $GaAs$, $GaSb$) is much smaller compared to that for hydrostatic strain. 

\section{Experimental Results}
\begin{figure*}[!ht]
\includegraphics[width=0.9\textwidth]{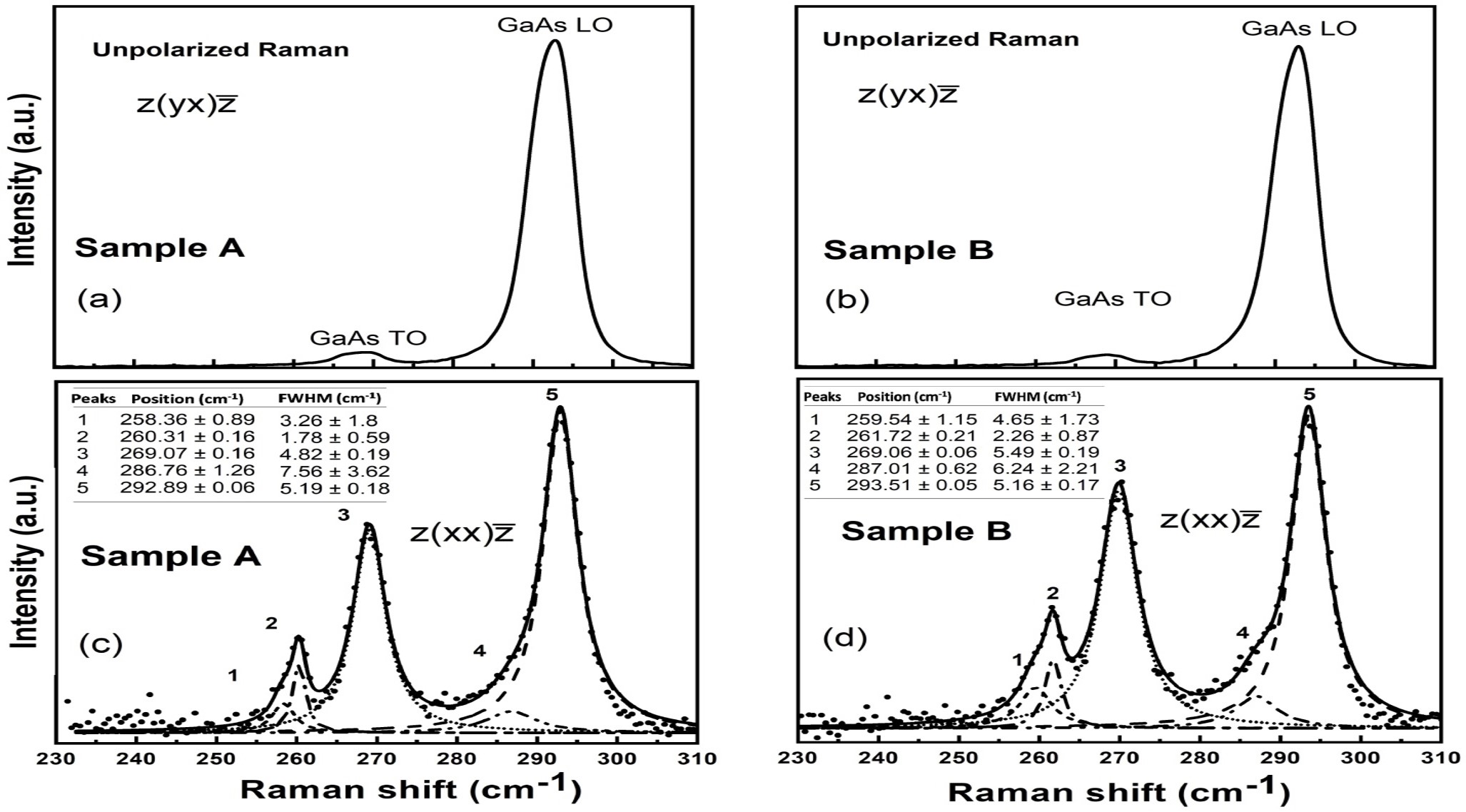}
\caption{\label{fig:interface_QD}Unpolarized Raman spectra of (a) sample A, $\&$ (b) sample B acquired in $z(yx)\overline{z}$ mode. Polarized Raman spectra of (c) sample A, $\&$ (d) sample B acquired in $z(xx)\overline{z}$ mode.}
\end{figure*}

The Raman selection criteria for backscattering geometry for $GaAs$ type $z[001]$ substrates predicts $LO$ allowed and $TO$ forbidden for $z(xy)\overline{z}$ geometry and $LO$ and $TO$ both forbidden for  $z(xx)\overline{z}$ geometry. Fig.~\ref{fig:interface_QD} (a) and (b) exhibit the Raman spectra acquired in $z(xy)\overline{z}$ geometry for samples A and B. It can be seen that only two distinct peaks are visible, which are attributed to $GaAs$ ($TO$) and ($LO$) phonons at $\approx270 cm^{-1}$  and $\approx294 cm^{-1}$  respectively. This is expected since the penetration depth of $532 nm$ laser line for GaAs is  $\approx150 nm$  and most of the signal arises from the top $100 nm$ $GaAs$ capping layer and the UQD with $GaAsSb$ capping layer, however, the contribution of the Raman signals arising from the LQD and $GaAsSb$ layers is relatively poorer. Further, it has been earlier reported that the $InAs$ and $GaAs$ like IF modes appear close to $LO$ peak positions and are very small in intensity\cite{pusep1998raman}, and thus the $z(xx)\overline{z}$ geometry is suitable to minimize the contributions for both the $GaAs$ $LO$ and $TO$ modes. It is known that the Raman linewidth reduces with decreasing temperature, parallel polarized Raman measurements in the later geometry were recorded at $77K$ to visualize different peaks originating from the corrugated interfaces of $InAs/GaAs$, $InAs/GaAsSb$, and $GaAsSb/GaAs$. Fig.~\ref{fig:interface_QD} (c) and (d) show low temperature ($77K$) Raman spectra of both the samples measured in $z(xx)\overline{z}$ scattering geometry and they were best fitted with five Lorentzian curves. The highest intensity peaks, peak 3 and peak 5 correspond to $GaAs$ $TO$ ($\approx269 cm^{-1})$, and $GaAs$ $LO$ ($\approx293 cm^{-1})$ phonon modes respectively. The other peaks correspond to $InAs$ QDs and $GaAs$, $GaAs_{1-x}Sb_x$ capping layer phonon modes. Also, the Raman peaks have a broad asymmetrical feature around $260 cm^{-1}$ (peak $2$) and $293 cm^{-1}$ (peak $5$). These modes are at slightly different frequency positions for both samples. The frequency positions are right-shifted to higher energy in sample B ($20\%$ $Sb$-content) as compared to sample A (having $10\%$ $Sb$-content). The higher $Sb$ content in sample B reduces the strain over the entire heterostructure (as earlier observed in the simulated strain profiles), which results in shift of the peak positions from their bulk value. The frequencies of $TO$ and $LO$ phonons in bulk $InAs$ are $218 cm^{-1}$ and $243 cm^{-1}$, respectively\cite{pusep1998raman}, whereas we observe two peaks at $260.31 cm^{-1}$ and $261.72 cm^{-1}$ in samples A and B respectively. These peak positions are at much higher frequencies than the $InAs$ modes but considerably lower than those for $GaAs$. The origin of these modes can be explained using our theoretical analysis for uniaxial and biaxial stress-induced shift of the optical phonon modes in QDs using the method outlined in the previous section (Theoretical Analysis) for both samples A and B. Fig.~\ref{fig:my_plot} (c) and (d) show that the $InAs$ $LO$ phonon modes are shifted to higher frequency $\approx260 cm^{-1}$, which is attributed to the compressive hydrostatic and tensile biaxial strain. Thus, we assign these modes (peak 2) as $InAs$ $LO$ peaks. This agrees with the results reported by other authors. Raman spectroscopic study by  Puesp et al\cite{pusep1998raman} on self-assembled QDs have observed the $InAs$ QD LO mode at $258 cm^{-1}$ which agrees very well with our value. Similarly, Milekhin et al\cite{milekhin2004raman} also assigned a value of $255 cm^{-1}$ for this mode.

In addition to the strong $LO$ and $TO$ modes of $GaAs$ (peaks $5$ and $3$) and $InAs$ QD $LO$ mode( peak $2$), there are additional weak shoulder peaks present at lower frequency side of $LO$ phonons (peak $4$ and peak $1$) which cannot be assigned to optical phonons. We ascribe them to be IF modes between the $InAs$ QD in $GaAs$ and $GaAsSb$ matrix. The frequency positions of the $GaAs$-like IF peak are at $\approx287 cm^{-1}$ (peak 4) in both the samples. This peak position is right shifted towards higher energy from the earlier reported\cite{milekhin2004interface} position of $283 cm^{-1}$ because of the inclusion of $Sb$ composition in the capping layer which causes strain relaxation in the heterostructure\cite{milekhin2004interface} with shift towards the bulk value. Another distinct peak at $\approx 259 cm^{-1}$ (peak $1$) is attributed to the IF mode of $InAs$ QD like $LO$ mode. This peak is at position $258.36 cm^{-1}$ and $259.54 cm^{-1}$ in samples A and B, respectively. Furthermore, the Full Width Half Maximum (FWHM) of most of the peaks (peak $1$, $2$ and $3$) is larger in sample B as compared to sample A. This might be due to the higher $Sb$-composition in the capping layer, which causes the localization of electrons in one material ($InAs$ QDs) and holes in another material ($GaAsSb$ capping region), thereby transforming the band alignment to be of type-II nature. However, all the peak positions in sample B are right-shifted by more than $1 cm^{-1}$, which is due to the bigger dot size and better strain relaxation inside the heterostructure (as discussed earlier). Dai et al\cite{dai2015raman} had attributed a mode at $230 cm^{-1}$ $GaAs_{1-x}Sb_x$ alloy mode with longest $Sb$ spray time, which we don’t observe due to poor S/N value over $220-250 cm^{-1}$. However a Raman study of $GaAs_{1-x}Sb_x$ with varying $x$ (Yano et al\cite{yano1989molecular} ) indicates that it can be assigned as $GaSb$ pure $TO$ mode. In addition, they assign $220 cm^{-1}$ as $InAs$ QD mode while we observe the $InAs$ QD $LO$ mode at $260 cm^{-1}$. We can estimate the frequency of the IF modes by following the prescription by Knipp and Reinecke\cite{knipp1992classical} based on the classical continuum dielectric approach without retardation, whereby the IF mode frequencies fall with the non-overlapping regions of the dot and the barrier materials. This is valid for $InAs$ QDs in $GaAs$ barrier and assuming the QDs as ellipsoidal shapes with height and width obtained from the cross-sectional TEM. We calculate the IF frequencies by following the code given by Hussay\cite{hussey2007raman} using Mathematica. The calculated frequencies for \textit{l=1-3} are shown in Table 1. According to the calculation, the first IF mode (peak 4) matches well with \textit{l=3} and \textit{m=0} in both samples, while IF mode with $l=1$ and $ m=0$ is closer to the strong $GaAs$ $LO$ mode. Thus, peak $1$ and peak $4$ are assigned as $InAs$-like IF and $GaAs$-like IF modes respectively.


\begin{table}[h]
\caption{Calculated IF modes of sample A and B.}
\centering
\resizebox{0.37\textwidth}{!}{%
\begin{tabular} {|c|c|c|c|}
\hline
\multirow{2}{*}{{Sample}} & $l=1$ & $l=2$ & $l=3$ \\ 
& $m = 0, 1$ & $m = 0, 1, 2$ & $m = 0, 1, 2, 3$ \\ \hline
\multirow{4}{*}{A}&  & $288.222$  & $286.917$ \\ 
& $290.193$ & $285.094$ & $285.156$ \\
& $284.678$ & $283.792$ & $284.022$ \\ 
&  &  & $283.525$ \\ \hline
\multirow{4}{*}{B}&  & $288.77$ & $287.432$ \\ 
& $290.789$ & $285.476$ & $285.553$ \\
& $285.038$ & $284.14$ & $284.377$ \\ 
& & & $283.872$ \\ \hline
\end{tabular}%
}
\label{table:gompertzfit}
\end{table}



 
\section{Conclusion}
In conclusion, we have identified distinct Raman peaks in the $InAs/GaAs_{1-x}Sb_x$ QD heterostructures. The phonon modes exhibited a shift to higher frequencies due to the increased $Sb$-composition in the capping layer. Additionally, the FWHM of the peaks was broader, indicating increased disorder and type-II band alignment. The QD heterostructure with a $GaAs_{0.8}Sb_{0.2}$ capping layer demonstrated a reduced strain profile, attributed to the lower lattice mismatch between $InAs$ QDs and the capping material. This reduction in strain led to larger quantum dots and resulted in longer wavelength emission. All observed modes have been assigned to the highly strained $LO$ mode of the $InAs$ QD and the $LO$ and $TO$ modes of $GaAs$. Additionally, we have identified two IF modes, which we have assigned as $GaAs$-like and $InAs$-like IF modes.

\begin{acknowledgments}
Financial support was provided by the Indian Institute of Technology Gandhinagar under grant number IP/ITGN/EE/JS/2122/08. The authors acknowledge Baolai Liang of UCLA for the growth of the samples and Sophisticated Analytical Instrument Facility) SAIF, IIT Bombay. 
\end{acknowledgments}
\nocite{*}
\bibliography{aipsamp}

\end{document}